\documentclass[10pt]{article}
\usepackage{caption}

\usepackage{subcaption}
\usepackage[utf8]{inputenc}

\usepackage{amssymb}
\usepackage{amsfonts}
\usepackage{amsmath}
\usepackage{amsthm}
\usepackage{mathtools}

\usepackage{graphicx}
\usepackage{indentfirst}
\usepackage{cite}
\usepackage{enumerate}
\usepackage{url}

\usepackage{lmodern}
\usepackage{newpxtext}

\usepackage[symbol]{footmisc}

\usepackage{tikz}
\definecolor{darkblue}{rgb}{0.15,0.35,0.55}
\definecolor{reddish}{rgb}{.8, 0.2, 0.2}

\usepackage[pdfpagelabels, linktocpage=true]{hyperref}
\hypersetup{
colorlinks=true,
citecolor=darkblue,
linkcolor=reddish,
urlcolor=darkblue,
pdfauthor={},
pdftitle={},
pdfsubject={}
}

\long\def\ca#1\cb{} 

\newcommand{\becs}{\begin{cases}}
\newcommand{\bem}{\begin{matrix}}

\newcommand{\dya}[1]{|#1\rangle\langle#1|}
\newcommand{\dyad}[2]{|#1\rangle\langle#2|}

\newcommand{\encs}{\end{cases}}
\newcommand{\enm}{\end{matrix}}

\newcommand{\ket}[1]{|#1\rangle }

\newcommand{\ot}{\otimes }

\newcommand{\Tr}{{\rm Tr}}

\newcommand{\BC}{{\mathcal B}}
\newcommand{\CC}{{\mathcal C}}
\newcommand{\DC}{{\mathcal D}}
\newcommand{\EC}{{\mathcal E}}

\newcommand{\HC}{{\mathcal H}}

\newcommand{\PC}{{\mathcal P}}
\newcommand{\QC}{{\mathcal Q}}

\newcommand{\TC}{{\mathcal T}}

\newcommand{\pB}{\textbf{p}}
\newcommand{\qB}{\textbf{q}}
\newcommand{\rB}{\textbf{r}}

\newcommand{\uB}{\textbf{u}}
\newcommand{\vB}{\textbf{v}}
\newcommand{\wB}{\textbf{w}}

\newcommand{\al}{\alpha }
\newcommand{\bt}{\beta }

\newcommand{\Gm}{\Gamma }
\newcommand{\dl}{\delta }
\newcommand{\Dl}{\Delta }
\newcommand{\ep}{\epsilon}

\newcommand{\zt}{\zeta }

\newcommand{\lm}{\lambda }
\newcommand{\Lm}{\Lambda }

\newcommand{\sg}{\sigma }

\theoremstyle{remark}
\newtheorem{theorem}{Theorem}
\newtheorem{corollary}{Corollary}

\usepackage{amsthm}
\theoremstyle{remark}
\newtheorem{lemma}{Lemma}

\usepackage[affil-it]{authblk}
\title{Allowing leakage can increase quantum transmission}
\author[1,2]{Vikesh Siddhu}
\affil[1]{Department of Physics \& Quantum Computing Group, Carnegie Mellon University}
\affil[2]{JILA, University of Colorado Boulder}
\date{}

\begin{document}
	\maketitle

\begin{abstract}
    Entanglement and quantum information lie at the root of quantum theory.
    These remarkable resources are generally believed to diminish when systems
    carrying them interact with their environment. By contrast, we find that
    engaging a system with its environment can increases its ability to carry
    quantum information.  The maximum rate of transmission is given by the
    quantum channel capacity. We counter-intuitively boost this capacity of a
    channel by allowing it to leak almost all quantum information to the
    channel's environment. 
    We conceptually and numerically explain this boost to arise from two-letter
    level super-additivity in the channel's coherent information. Such
    super-additivity has a far larger magnitude and a qualitatively wider
    extent than previously known. Our findings have a surprising implication
    for quantum key distribution: maximum rates for key distribution can be
    boosted by allowing leakage of information to the eavesdropping
    environment. 
\end{abstract}

\tableofcontents

\section{Introduction}
\label{sec:intro}

Quantum information is an extremely useful resource for computation and
communication~\cite{NielsenChuang11}. However understanding this useful
resource remains a challenge: quantum information has no precise agreed upon
definition and it can even be viewed to become
negative~\cite{HorodeckiOppenheimEA05}. Whatever this quantity is, it cannot be
copied without being disturbed~\cite{WoottersZurek82, Park70, Dieks82}. This
no-cloning property makes quantum information even more susceptible to noise
than it would ordinarily be due to the delicate nature of quantum systems that
carry this information.
How noise wilts away quantum information, and how much of it can be kept
noiseless are key questions~\cite{BennettShor98}. These fundamental questions
inform practical efforts to mitigate and fully correct for noise in
applications of quantum computation and communication~\cite{RenesDupuisEA12,
WildeGuha13, LeverrierZemor22, BravyiCrossEA23, PanteleevKalachev22,
GoswamiMhallaEA23}.

The key question we ask is, given a simple noise model, at what maximum rate
$\QC$~(called the quantum capacity) can quantum information pass perfectly as
noisy qubits carrying this information are scaled asymptotically.  If such a
noise model is modified to allow for additional leakage of quantum information,
then does the modification increase or decrease the quantum capacity $\QC$?
Conventional wisdom on such leakage is as follows. Since the no-cloning
principle prevents quantum information from noiselessly appearing in more than
one place, the part of the information accessible to the environment gets lost,
i.e., leaked away from the output. Thus allowing leakage of information to the
environment is not expected to improve one's ability to keep quantum
information noise free.

Our key theoretical discovery is a simple, physically grounded noise model
(also called a quantum channel) whose quantum and private capacities can be
computed, but for which we prove how a counter-intuitive scheme to allow
leaking of almost all quantum information to the channel's environment boosts
the channel's quantum and private capacities. We argue the origin of this boost
to be super-additivity~\cite{SmithYard08}, a fundamentally quantum synergy
between point-to-point quantum channels absent from its direct classical
analog~\cite{Shannon48}. 
Taming super-additivity is one of the longest-standing fundamental problems in
quantum information theory~\cite{DiVincenzoShorEA98, DevetakShor05,
BradlerDutilEA10, Smith10, Watanabe12, ElkoussStrelchuk15, WangDuan16,
SutterScholzEA17, LimLee18, LimTakagiEA19, FanizzaKianvashEA20,
NohPirandolaEA20, KhatriSharmaEA20, ChessaGiovannetti21}.
The extent, magnitude, and origin of super-additivity here is dramatically
wider, larger, and novel, respectively.

Before our technical discussion we make a simple conceptual observation about
our result and quantum entanglement~\cite{HorodeckiHorodeckiEA09a}.
Entanglement is tightly linked with quantum information via quantum
teleportation~\cite{BennettBrassardEA93}.
Passing noiseless quantum information across a noisy system is equivalent to
preserving entanglement of that system with a noiseless system and leaking is
equivalent to becoming entangled with an unwanted~(environment)
system~\cite{BarnumKnillEA00, KretschmannWerner04, Wilde17}.
Because entanglement is monogamous~\cite{CoffmanKunduEA00,
Terhal04}~(roughly speaking an increase in entanglement between a pair of
systems accompanies a decrease in entanglement of the pair with any third
system), and one expects quantum systems that become entangled with their
environment to become worse at retaining entanglement with a reference system.
Thus, allowing an increase in the ability to entangle with a channel’s
environment is not expected to increase the channel’s quantum capacity. 
We not only prove that such an increase arises in a channel presented here but
also argue the increase appears due to super-additivity.

The rest of this manuscript is organized as follows. Our notation, definitions
of quantum capacities, along with a technical tool analyze them are covered in
Sec.~\ref{sec:prelim}. The next section, Sec.~\ref{sec:qbitChan}, introduces
the isometry for our simple qubit input channel and disucsses its properties.
Sec.~\ref{sec:QtritInpCh} has a dicussion of the extended channel obtained by
allowing leakage. Super-additivity in this channel's coherent information is
analyzed in Sec.~\ref{sec:superAdd}. The next section, sec.~\ref{sec:QCapRel}
is devoted to a discussion comparing the capacities of the original and extend
channel. The final section, Sec.~\ref{sec:disc}, contains a dicussion of the
results. Two appendices, App.~\ref{AQutritCoh2} and~\ref{AQtritEnt}, contain
various technical details.

\section{Preliminaries}
\label{sec:prelim}

\subsection{Notation and quantum channels}

Let $\HC$ represent a finite dimensional Hilbert space with dimension $d$.
Let $\{ \ket{i}\}$ represent the standard basis of $\HC$ and
$\hat \HC$ represent the space of linear operators on $\HC$. A positive
semi-definite operator $\rho \in \hat \HC$ with unit trace, $\Tr(\rho) = 1$, is called a
density operator. The support of a density operator is the subspace spanned
by its eigenvectors with non-zero eigenvalues.
When $d=2$, $\rho$ can be written as

\begin{equation}
    \rho = \frac{1}{2} \big( I + x X + y Y + z Z \big),
\end{equation}
where $\rB:= (x,y,z)$ is called a Bloch vector, satisfying $\rB \cdot \rB \leq
1$, and $X = \dyad{0}{1} + \dyad{1}{0}$, $Y = i\dyad{0}{1} -i \dyad{1}{0}$, and $Z
= \dya{0} - \dya{1}$ are Pauli matrices.
We use

\begin{equation}
    S(\rho) := -\Tr (\rho \log \rho),
    \label{eq:vonNeu}
\end{equation}
to denote the von-Neumann entropy of $\rho$~(all $\log$ functions are in base 2).

An isolated quantum system with Hilbert space $\HC_a$ becomes noisy by
interacting with its environment.  Such an interaction is captured by an
isometry,

\begin{equation}
    J : \HC_{a} \mapsto \HC_{b} \ot \HC_{c}; \quad J^{\dag} J  = I_a,
    \label{isoBasic}
\end{equation}
taking a system's Hilbert space $\HC_a$ to a subspace of the joint
output-environment space $\HC_b \ot \HC_c$, where $\ot$ represents tensor
product.
Sometimes we use a more compressed notation $J:a \mapsto b \ot c$.
The interaction given by $J$ maps any system
density operator $\rho_a$ to a pair of density operators

\begin{equation}
    \rho_b = \BC(\rho_a) := \Tr_c(J \rho_a J^{\dag}), \quad
    \text{and} \quad 
    \rho_c = \CC(\rho_a) := \Tr_b(J \rho_a J^{\dag}),
    \label{pairDef}
\end{equation}
at the output and environment, respectively, here $\Tr_c$ represents partial
trace over $c$ and $\Tr_b$ is defined similarly.  In this way $J$ defines two
completely positive trace preserving maps, i.e., quantum channels, $\BC: \hat
\HC_a \mapsto \hat \HC_b$ and $\CC: \hat \HC_a \mapsto \hat \HC_c$.
The quantum channel $\CC$ to the system's environment is called the complement
of $\BC$. 
Restricting the input of $J$ to a subspace $\HC_{a'}$ of $\HC_a$ results in an
isometry from $\HC_{a'}$ to $\HC_b \ot \HC_c$. Channels, $\BC' : \hat
\HC_{a'} \mapsto \hat \HC_b$ and $\CC': \hat \HC_{a'} \mapsto \hat \HC_c$,
defined by this isometry may be referred to as sub-channels of $\BC$ and $\CC$,
respectively.

Let $\{ \ket{i}_c \}$ and $\{ \ket{j}_b \}$ be orthonormal basis of $\HC_c$ and $\HC_b$
respectively, then $J = \sum_i L_i \ot \ket{i}_c = \sum_j K_j \ot \ket{j}_b$
where $\{L_i\}$ and $\{K_j\}$ are called Kraus operators for the $\BC$
and $\CC$ channels, respectively, i.e.,

\begin{equation}
    \BC(A) = \sum_i L_i A L_i^{\dag} \quad \text{and} \quad
    \CC(A) = \sum_j K_j A K_j^{\dag}.
\end{equation}

\subsection{Channel capacities}

The {\em coherent information} of $\BC$ at $\rho_a$,

\begin{equation}
    \Dl(\BC ,\rho_a) = S(\rho_b) - S(\rho_c),
    \label{entBias}
\end{equation}
can be maximized over all input density operators to obtain the {\em channel
coherent information}~(also called the one-letter coherent information),

\begin{equation}
    I(\BC) = \max_{\rho_a} \Dl(\BC, \rho_a).
    \label{chanCoh}
\end{equation}
A channel's coherent information is an achievable rate for sending quantum
information across the channel. The largest possible achievable rate defines a
channel's quantum capacity. For any channel $\BC$ this quantum capacity is
given by a multi-letter formula~\cite{Lloyd97, BarnumKnillEA00, Shor02a, Devetak05},

\begin{equation}
    \QC(\BC) = \lim_{n \mapsto \infty} \frac{1}{n} I(\BC^{\ot n}),
    \label{chanCap}
\end{equation}
where $\BC^{\ot n}$ represents $n \in \mathbb{N}$ parallel uses of
$\BC$. The limit in eq.~\eqref{chanCap} is intractable to compute due to
super-additivity of $I$: for two channels $\BC$ and $\BC'$ used in
parallel, the inequality,

\begin{equation}
    I(\BC \ot \BC') \geq I(\BC) + I(\BC'),
    \label{nonAddDef}
\end{equation}
can be strict. There are exotic examples of channels which display
super-additivity: for any integer $l>1$ there is a channel $\tilde \BC$ for which
$I(\tilde \BC^{\ot l}) = 0$ but $I(\tilde \BC^{\ot k})>0$ for
some $k>l$~\cite{CubittElkoussEA15}. Such examples show that
using eq.~\eqref{chanCap} to check positivity of $\QC$ is hard. 

The private capacity $\PC$ is the analog of the quantum capacity $\QC$ for
sending classical information across a channel that is hidden from the channel's
environment. Like $\QC$, $\PC$ is given by a multi-letter
formula of the form eq.~\eqref{chanCap}, where $I$ is now replaced with the
{\em private information},

\begin{equation}
    I_p(\BC) = \max_{ \{p(x), \rho_a(x)\}} \Dl(\BC,\rho_a) - \sum_x p(x)
    \Dl\big(\BC,\rho_a(x)\big),
    \label{pInfDef}
\end{equation}
where $\rho_a = \sum_x p(x) \rho_a(x)$. The private information is
super-additive in exactly the same way as $I$ in eq.~\eqref{nonAddDef}. The
maximization in eq.~\eqref{pInfDef} gives $I(\BC)$ when every $\rho_a(x)$ is a
pure state, hence, $I(\BC) \leq I_p(\BC)$ and thus $\QC(\BC) \leq \PC(\BC)$.
Both these inequalities can be strict~\cite{LeungLiEA14, LeditzkyLeungEA18}.
The strictness can be extreme in the sense that a channel with zero quantum
capacity can have strictly positive private
capacity~\cite{HorodeckiHorodeckiEA05,HorodeckiPankowskiEA08,HorodeckiHorodeckiEA09}. 

Equality, $\QC(\BC) = \PC(\BC)$, is known to hold when $\BC$'s complement $\CC$
has zero quantum capacity, i.e., $\QC(\CC) = 0$. When $\PC(\CC)$ is also zero,
additivity holds in the sense that~\cite{Watanabe12},
\begin{equation}
    I(\BC) = \QC(\BC) = \PC(\BC) = I_p(\BC).
    \label{qpEqual}
\end{equation}
This additivity greatly simplifies the discussion of the quantum and private
capacities of $\BC$. However, this simplicity comes with the headache of
showing that a channel's complement has no quantum and no private capacity.
For a general channel this headache has no known cure due to super-additivity.
Under special circumstances when $\BC$ is {\em degradable} and thus its
complement $\CC$ {\em anti-degradable}~\cite{DevetakShor05}, i.e., $\DC \circ
\BC = \CC$ for some quantum channel $\DC$, one can show that $\QC(\CC) =
\PC(\CC) = 0$~\cite{LeditzkyDattaEA18}. Thus, degradability
implies~\eqref{qpEqual}, but in addition it enforces a more general additivity:
for any two channels $\BC$ and $\BC'$, either degradable or anti-degradable,
equality holds in eq.~\eqref{nonAddDef}. Due to their pleasant properties,
(anti)degradable channels have also been widely
studied~\cite{CarusoGiovannetti06, CubittRuskaiEA08, BradlerDutilEA10}. Such
studies show that (anti)degradability can be checked using a semi-definite
program~\cite{SutterScholzEA17} and used, along with other methods~\cite{
    HolevoWerner01, MullerHermesReebEA16, WangDuan16, WangFangEA18}, to find
computable bounds on quantum capacities~\cite{SmithSmolin08,Ouyang14,
SutterScholzEA17, LeditzkyDattaEA18,FanizzaKianvashEA20}.

\subsection{Log-singularity}
We review an algebraic technique to analyze a channel's coherent 
information~\cite{Siddhu21}~(see also~\cite{SinghDatta22}).
Let $\rho(\ep)$ be a density operator which depends on a real parameter $\ep$
and $S(\ep):= -\Tr\big( \rho(\ep) \log \rho(\ep) \big)$ denote its von-Neumann
entropy~($\log$ is base 2).  If one or several eigenvalues of $\rho(\ep)$
increase linearly from zero to leading order in $\ep$, then we say $S(\ep)$
has an {\em $\ep \log$ singularity}. Due to this so-called
singularity, $S(\ep) \simeq x |\ep \log \ep|$ for small $\ep$, where $x>0$ is
referred to as the {\em rate} of this $\ep \log$-singularity. 

A channel $\BC$ with complement $\CC$ maps an input density operator
$\rho_a(\ep)$ to an output $\rho_b(\ep) := \BC\big(\rho_a(\ep)\big)$ and
complementary output $\rho_c(\ep):= \CC \big( \rho_a(\ep)\big)$.  Denote the
coherent information of $\BC$ at $\rho_a(\ep)$ by $\Dl(\ep) := S_b(\ep) -
S_c(\ep)$, where $S_b(\ep):= S\big(\rho_b(\ep)\big)$ and $S_c(\ep):=
S\big(\rho_c(\ep)\big)$. If an $\ep \log$ singularity is present in either one
or both $S_b(\ep)$ and $S_c(\ep)$, then the singularity with larger rate is
said to be {\em stronger}.

A $\log$-singularity based argument to show $I(\BC)>0$ proceeds as
follows: choose $\rho_a(\ep)$ such that at $\ep = 0$, $\Dl(0) = 0$ and
$S_b(\ep)$ has a stronger $\ep \log$ singularity than $S_c(\ep)$. Then for
small $\ep$, $\Dl(\ep) \simeq |O(\ep \log \ep)|>0$. Since $\Dl(\ep) \leq
I(\BC)$, we have a non-zero $I(\BC)$.
A similar argument to show super-additivity,

\begin{equation}
    I(\BC \ot \BC) > 2 I(\BC),
\end{equation}
proceeds as follows. For an input $\rho_{aa}(\ep)$ to $\BC \ot \BC$ denote the
entropy of the output $\rho_{bb}(\ep)$ and the complementary output
$\rho_{cc}(\ep)$ by $S_{bb}(\ep)$ and $S_{cc}(\ep)$, respectively. Choose input
$\rho_{aa}(\ep)$ 
such that at $\ep = 0$, $\Dl(0) = 2I(\BC)$ and $S_{bb}(\ep)$ has a stronger
$\ep \log$ singularity than $S_{cc}(\ep)$. Then, for small $\ep$ $\Dl(\ep) -
\Dl(0) \simeq |O(\ep \log \ep)| > 0$. Since $I(\BC \ot \BC) \geq \Dl(\ep)$, we
have $I(\BC \ot \BC)> 2I(\BC)$.

\section{A qubit input channel $\BC$}
\label{sec:qbitChan}

\begin{figure}[ht]
    \centering
    \begin{subfigure}[b]{0.45\textwidth}
        \centering
        \includegraphics[scale=0.9]{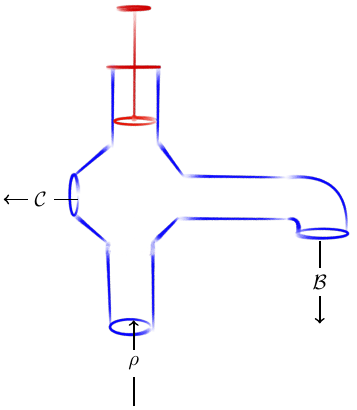}
        \caption{Here an input quantum state $\rho$~(enters from below) is
        transformed to a joint output-environment state, the environment state
        $\CC(\rho)$~(leaks behind) and output $\BC(\rho)$~(moves forward)
        separately depart to their respective spaces.}
    \label{FigA1}
    \end{subfigure}
    \hfill
    \begin{subfigure}[b]{0.45\textwidth}
        \centering
        \includegraphics[scale=1.0]{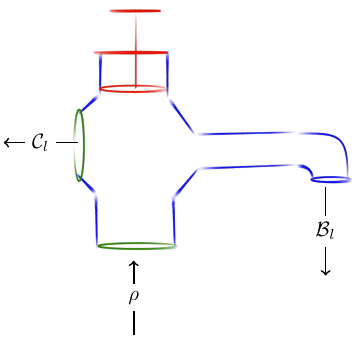}
        \caption{The leakage procedure applied to Fig.~\ref{FigA1} expands the
        input space~(circle below is stretched) and the environment
        space~(circle left is stretched) such that any input in the expansion
        leaks perfectly to the environment while the output space~(circle
        right) remains unchanged.}    
    \label{FigA2}
    \end{subfigure}
    \caption{Enhancing quantum transmission by allowing leakage: schematic
    diagrams representing a quantum channel pair~(a) and its leaked
    version~(b).}
\end{figure}

Consider a qubit system ground state $\ket{0}$ remains intact, however its
first excited state $\ket{1}$ is noisy.
Due to noise, $\ket{1}$ decays to the ground state
with probability $\lm$, it remains in the $\ket{1}$ state with probability
$(1-\lm)/2$, otherwise it leaks out the $\{\ket{0}, \ket{1}\}$ qubit subspace
to an excited state, say $\ket{2}$.
Model this noise mathematically modelled using an isometry, $J: a \mapsto b \ot
c$, 
\begin{equation}
    J \ket{0} = \ket{01}, \quad 
    J \ket{1} = \sqrt{\lm} \ket{00} + \sqrt{\frac{1-\lm}{2}} \ket{11} + \sqrt{\frac{1-\lm}{2}} \ket{22},
    \label{eq:iso1}
\end{equation}
mapping an input qubit system $\rho_a$ to a qutrit output $\BC(\rho_a)$
and qutrit environment $\CC(\rho_a)$~(see Fig.~\ref{FigA1}).

\subsection{Damping and leakage property}
\label{sec:AAD}
Channel $\BC$ has a rich structure, it represents both damping and leakage
present in physical realizations of qubits~\cite{MehlkopfKorbeeEA84,
AliferisTerhal07, SahayJinEA23, MiaoMcEwenEA23, ChirolliBurkard08,
HouckSchreierEA08}.  In such realizations, if the system escapes the
$\{\ket{0}, \ket{1} \}$ subspace then one may post-select.  Done carefully, for
instance by detecting leakage to $\ket{2}$ and resetting the system to
$\ket{1}$, this post-selection results in a qubit to qubit noise channel that
damps $\ket{1}$ to $\ket{0}$ and dephases the coherence between the two
states.
Let $P_b = [0] + [1]$ be a projector onto the $\{\ket{0}, \ket{1}\}$ qubit
subspace of $\HC_b$, then post-selection to detect leakage and reset
the system to $\ket{1}$ results in a quantum channel,
\begin{equation}
    \BC'(A) = P_b \BC(A) P_b + \dyad{1}{2} \BC(A) \dyad{2}{1}.
\end{equation}
One can show that such post-selection satisfies,
\begin{equation}
    \BC'(A) = P_b \BC'(A) P_b,
\end{equation}
i.e., the output is always in the $\{ \ket{0}, \ket{1}\}$ subspace. This operator
can thus be adequately represented using a $2 \times 2$ matrix obtained by deleting
the third row and column of $\BC'(A)$. When the input
$A$ is a density operator with Bloch vector $\rB = (x, y, z)$ then the output
of $\BC'$ has a Bloch vector

\begin{equation}
    \rB' = \big( \sqrt{\frac{1 - \lm}{2}} x,  \sqrt{\frac{1 - \lm}{2}} y,
    \lm + z (1-\lm) \big).
\end{equation}
Note the $z = 1$ state, $\dya{0}$, is mapped to itself, the $z = -1$ state, 
$\dya{1}$, damps to $\dya{0}$ with probability $\lm$. Coherence between
$\ket{0}$ and $\ket{1}$, represented by first two entries of the Bloch vector
shrink by a factor of $\sqrt{(1-\lm)/2}$ under the action of $\BC'$.
%

\subsection{(Anti)~Degradability}
\label{AADeg2}

To fully understand the flow of information in a noise model described by some
isometry $J$, we focus on both channels that $J$ generates: $\BC$ to the output
and $\CC$ to the environment. The channel to the output can simulate the
channel to the environment or be simulated by it, as expressed by this next
Lemma.

\begin{lemma}: 
Channel $\BC$ is degradable for $0 \leq \lm \leq 1/3$ and anti-degradable
    for $1/3 \leq \lm \leq 1$.  
    \label{lem:degADeg}
\end{lemma}
\begin{proof}
To show that $\BC$ is degradable for $0 \leq \lm \leq 1/3$, we construct an
isometry $K: \HC_{b} \mapsto \HC_{c} \ot \HC_{d}$ of the form
\begin{equation}
    K \ket{0} = \ket{10}, \quad
    K \ket{1} = \sqrt{\dl}\ket{00} + \sqrt{1-\dl} \ket{11}, \quad
    K \ket{2} = \ket{22},
    \label{degIsoEq}
\end{equation}
where $\HC_{d}$ has dimension three and $0 \leq \dl \leq 1$. For $0 \leq \lm
\leq 1/3$, set $\dl = 2 \lm/(1-\lm)$ in the above equation; for any
operator $O \in \hat \HC_{b}$, the channel
\begin{equation}
    \DC(O) = \Tr_{d}(K O K^{\dag}),
    \label{degChan}
\end{equation}
satisfies
\begin{equation}
    \DC \circ \BC = \CC,
    \label{degCondQbit}
\end{equation}
i.e., $\BC$ is degradable.  
To show $\BC$ is anti-degradable for $1/3 \leq \lm \leq 1$ we use an isometry
$L : \HC_{c} \mapsto \HC_{b} \ot \HC_{e}$,
\begin{equation}
    L \ket{0} = \sqrt{1-\eta}\ket{00} + \sqrt{\eta} \ket{11}, \quad
    L \ket{1} = \ket{01}, \quad
    L \ket{2} = \ket{22},
\end{equation}
where $\HC_{e}$ is three dimensional and $0 \leq \eta \leq 1$. In the
parameter interval $1/3 \leq \lm \leq 1$, set $\eta = (1-\lm)/2\lm$. In
this setting, the isometry $L$ defines~(using an equation analogous to
\eqref{degChan}) a channel $\EC : \hat \HC_{c} \mapsto \hat \HC_{b}$ which
satisfies
\begin{equation}
    \EC \circ \CC = \BC,
    \label{aDegCondQbit}
\end{equation}
        i.e., $\BC$ is anti-degradable.
By definition, if $\BC$ is anti-degradable then $\CC$ is degradable, and if
$\BC$ is degradable then $\CC$ is anti-degradable. From the discussion
above, it follows that $\CC$ is anti-degradable for $0 \leq \lm \leq 1/3$
and degradable for $1/3 \leq \lm \leq 1$.  
\end{proof}

\subsection{Quantum capacity $\QC$ and coherent information $I$}
\label{AACohInfo}

Unlike most channel pairs, the quantum capacity
$\QC$ of both $\BC$ and $\CC$ can be found exactly; $\QC$ for each channel can
be shown to equal the channel's coherent information $I$ and this coherent
information too can be found with relative ease. This coherent information remains
generically positive, as shown next.

\begin{lemma}
For $0 \leq \lm \leq 1/3$ the coherent information 
    $\Dl\big(\BC, \sg_a(z) \big) \geq 0$ where $\sg_z(a) = z \dya{0} + (1-z) \dya{1}$ and $0 \leq z \leq 1$.
\end{lemma}
\begin{proof}
    For any $n$-dimensional real valued vector $\uB$ let $\uB^{\downarrow}$
    denote the vector with entries of $\uB$ arranged in descending order. 
    Use $\uB \prec \vB$ to denote $\uB$ is majorized by $\vB$, i.e., the inequality,
    \begin{align}
        \sum_{i=1}^{k} \uB_i^{\downarrow} \leq \sum_{i=1}^{k} \vB_j^{\downarrow},
    \end{align}
    holds for all $k \leq n$ and becomes an equality at $k=n$. Let $\pB$
    be a probability vector and $h(\pB):= \sum_i \pB_i \log \pB_i$ denote
    its Shannon entropy. The Shannon entropy is Schur concave, i.e., for
    two probability vectors $\pB$ and $\qB$ that satisfy
    $\pB \prec \qB$ the Shannon entropy satisfies $h(\pB) \geq h(\qB)$.
    For any density operator $\Lm$ we use the notation $\wB(\Lm)$ to denote
    its eigenvalues and $S(\Lm) = h(\wB(\Lm))$.
    Notice 
    \begin{equation}
        \sg_b:= \BC(\sg_a) = (1-z) \Gm + z \dya{0}, \quad \text{and} \quad
        \sg_c:= \CC(\sg_a) = (1-z) \Gm + z \dya{1},
    \end{equation}
    where $\Gm = \lm \dya{0} + (1-\lm)(\dya{1} + \dya{2})/2$ and $0 \leq z
    \leq 1$.  From Ex.II.1.15 in~\cite{Bhatia97} it follows that
    \begin{align}
        \begin{aligned}
            \wB(\sg_b)) & \prec (1-z) \wB^{\downarrow}(\Gm) + z \wB^{\downarrow}(\dya{0}) \\
            & =  \wB(\sg_c).
        \end{aligned}
    \end{align}
    Where the last equality uses $0 \leq z \leq 1$ and $0 \leq \lm \leq
    1/3$.  From the above equation, together with Schur-concavity of the
    Shannon entropy we find $h(\wB(\sg_b)) \geq h(\wB(\sg_c))$, i.e.,
    $S(\sg_b) \geq S(\sg_c)$.
\end{proof}

One can always compute $I(\BC)$ by maximizing the coherent information
$\Dl(\BC, \rho_{a})$ over density operators $\rho_{a}$ that are diagonal in the
standard basis.

\begin{figure}[ht]
    \centering
    \includegraphics[scale=0.75]{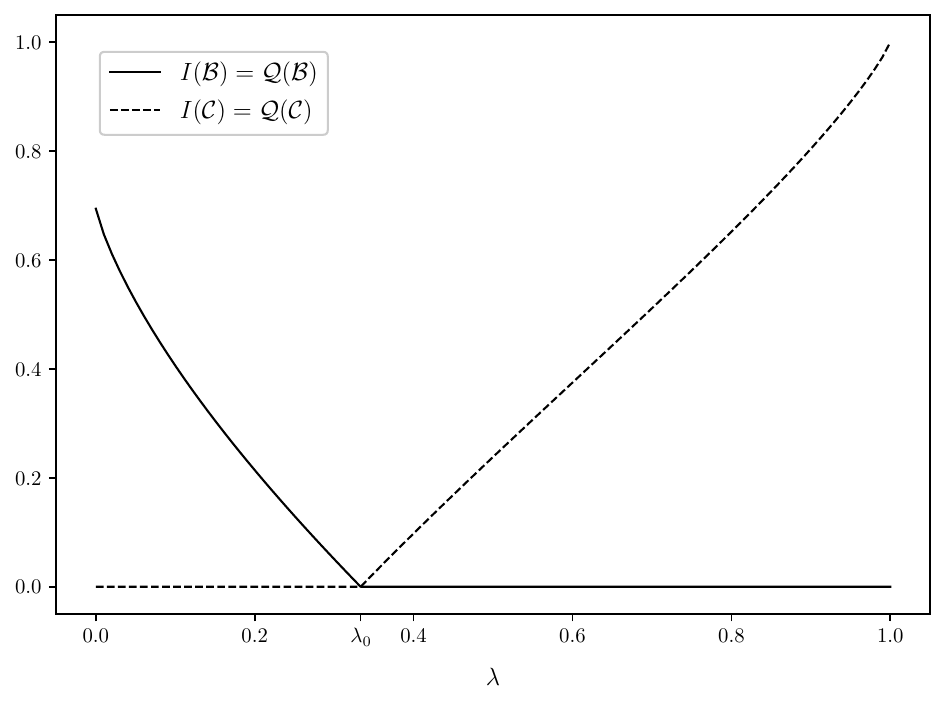}
    \caption{A plot of $I(\BC)$ and $I(\CC)$ versus $\lm$.
    The solid line shows $I(\BC) = \QC(\BC)$, it is maximum at
    $\lm = 0$ and decreases monotonically with $\lm$ to become zero at $\lm=
    1/3$. The dashed line shows $I(\CC) = \QC(\CC)$, it
    increases monotonically with $\lm$. The line has a constant value of
    zero for $0 \leq \lm \leq 1/3$, and becomes positive thereafter.}
    \label{FigE}
\end{figure}

    \begin{theorem}
        \label{th:QCapTh}
    For all $0 \leq \lm \leq 1$, the channel coherent information,
    \begin{equation}
        \QC(\BC) = I(\BC) = \underset{0 \leq z \leq 1}{\max} \Dl \big( \BC, \sg_{a}(z) \big ),
        \label{lm3:eq1}
    \end{equation}
        and
    \begin{equation}
        \QC(\CC) = I(\CC) = \underset{0 \leq z \leq 1}{\max} \Dl \big( \CC, \sg_{a}(z) \big ),
        \label{lm3:eq2}
    \end{equation}
    where $\sg_a(z) = z \dya{0} + (1-z) \dya{1}$.
    \end{theorem}
    \begin{proof}
        As $\BC$ is either degradable or anti-degradable~(see
        Lemma~\ref{lem:degADeg}), $I(\BC) = \QC(\BC)$ and $\QC(\CC) = I(\CC)$.
        We prove the second equality in~\eqref{lm3:eq1}. The proof for the
        second equality in~\eqref{lm3:eq2} is analogous.
    Consider unitary operators
\begin{equation}
    U_{a} = [0] - [1], \quad V_{b} = [0] - [1] + [2], \quad \text{and}
    \quad W_{c} = [0] - [1] + [2],
    \label{unitaryDef}
\end{equation}
on $\HC_{a}$, $\HC_{b}$, and $\HC_{c}$ respectively, where $[\psi]$ is our
notation for $\dya{\psi}$. The isometry $J$ in eq.~\eqref{eq:iso1} has a
symmetry in the sense that,
\begin{equation}
    J U_{a} = (V_{b} \ot W_{c}) J.
    \label{AQIsoSymm}
\end{equation}
In light of the above symmetry, the superoperators of the two channels
$\BC$ and $\CC$ defined by this isometry $J$ satisfy
\begin{equation}
    \BC(U_{a} A U_{a}^\dag) = V_{b} A V_{b}^{\dag}, \quad
    \CC(U_{a} A U_{a}^\dag) = W_{c} A W_{c}^{\dag},
    \label{AQChanSymm}
\end{equation}
where $A \in \hat \HC_a$. As a consequence of eq.~\eqref{AQChanSymm} the coherent
information of $\BC$ at any input density operator $\rho_{a}$ satisfies,
\begin{equation}
    \Dl(\BC, \rho_{a}) = \Dl(\BC, U_{a} \rho_{a} U_{a}^{\dag}).
     \label{AentBiasSymm}
\end{equation}
For $0 \leq \lm \leq 1/3$, the $\BC$ channel is degradable and thus
$\Dl(\BC,\rho_{a})$ is a concave function of $\rho_{a}$~\cite{YardHaydenEA08}.
Maximizing this concave function over all possible density operators $\rho_{a}$
gives $I(\BC)$.  Given any $\rho_{a}$, construct  
\begin{equation}
    \sg_{a} = \frac{1}{2} \big( \rho_{a} + U_{a} \rho_{a} U_{a}^{\dag} \big),
    \label{AsgDef}
\end{equation}
and observe
\begin{equation}
    \Dl(\BC, \sg_{a}) \geq \frac{1}{2} \big( \Dl(\BC, \rho_{a}) + 
    \Dl(\BC, U_{a} \rho_{a} U_{a}^{\dag}) \big) = \Dl(\BC, \rho_{a}),
    \label{AinEqCoh}
\end{equation}
where the inequality above follows from concavity of $\Dl(\BC, \rho_{a})$,
and the equality comes from eq.~\eqref{AentBiasSymm}. From eq.~\eqref{AsgDef} it
follows that $\sg_{a}$ is diagonal in the standard basis, i.e.,
\begin{equation}
    \sg_{a}(z) = z [0] + (1-z) [1],
    \label{sgOpt}
\end{equation}
where $0 \leq z \leq 1$. In the interval $0 \leq \lm \leq 1/3$,
eq.~\eqref{AinEqCoh} holds and thus
\begin{equation}
    I(\BC) = \max_{ 0 \leq z \leq 1 }\Dl \big(\BC, \sg_{a}(z) \big).
\end{equation} 
For parameter values $1/3 \leq \lm \leq 1$, $\BC$ is anti-degradable
and $I(\BC)=0$. Since $\Dl(\BC,\sg_{a}) = 0$ at $z = 0$, the
above equation holds for all $0 \leq \lm \leq 1$.
\end{proof}
Carrying out the optimization in eqns.~\eqref{lm3:eq1} and~\eqref{lm3:eq2} we
obtain $\QC(\BC)$ and $\QC(\CC)$, respectively. These values are plotted in
Fig.~\ref{FigE}. The plots that
$\QC(\BC)$ decreases monotonically with $\lm$, it is positive for $0 \leq
\lm < 1/3$ and zero for $1/3 \leq \lm \leq 1$ while
$\QC(\CC)$ increases monotonically with $\lm$, it is zero for $0 \leq
\lm \leq 1/3$ and strictly positive for $1/3 < \lm \leq 1$.

\section{Extended channel $\BC_l$}
\label{sec:QtritInpCh}

Our next step is to include higher levels $\ket{i}$, $i \geq 1$ in the input
system $a$ which damp to $\ket{0}$ with probability 1. The inclusion
modifies~\eqref{eq:iso1} into an isometry,

\begin{equation}
    J_l\ket{i} = 
    \begin{cases} 
        J \ket{i} & \text{if} \; i \leq 1 \\
        \ket{0i} & \text{otherwise}
    \end{cases}
    \label{eq:iso2}
\end{equation}
whose input system $al$ and environment system $cl$ have fixed dimension $d
> 2$ while the output system is unchanged and thus still called $b$~(see
Fig.~\ref{FigA2}). The isometry generates channels $\BC_l :al \mapsto b $ and
$\CC_l: al \mapsto cl$.
It allows leakage of almost all quantum information to the environment system.
This leakage comes across as an increase in the capacity of the channel to the
environment, $\QC(\CC_l) \geq \log (d-1)$ for all $\lm$, in contrast to
$\QC(\CC)$ which is zero for $0 \leq \lm \leq 1/3$.  In this $0 \leq \lm \leq
1/3$ parameter range, the previous channel $\CC$ could be simulated from its
complementary output $\BC$, however $\CC_l$ cannot be simulated the same way by
its complementary output $\BC_l$ for any $\lm$.
This is expected, the environment now has a lot of information the output
doesn't.  On the other hand, the environment can simulate the output for $\lm
\geq 1/2$ where $\BC_l$ is anti-degradable and $\QC(\BC_l) = 0$. We present
formal proofs in what follows.

\subsection{(Anti)~Degradability}

\label{AQutritSub}
Let $\HC_{a'}, \HC_{b'},$ and $\HC_{c'}$ denote subspaces of $\HC_{al}, \HC_{b},$
and $\HC_{cl}$ respectively, where each subspace is the support of

\begin{equation}
    P_{a'} = [0] + \sum_{i=2}^{d-1} [i], \quad
    P_{b'} = [0] , \quad \text{and} \quad
    P_{c'} = \sum_{i=1}^{d-1} [i],
\end{equation}
respectively. Clearly, both $\HC_{a'}$ and $\HC_{c'}$ have dimension $d-1$
while $\HC_{b'}$ is one dimensional.
Restricting the input of the isometry $J:\HC_{al} \mapsto \HC_{b} \ot \HC_{cl}$ to
the subspace $\HC_{a'}$ yields 

\begin{equation}
    J' := J_lP_{a'},\quad \text{where} \quad 
    (J')^{\dag} J' = P_{a'}.
\end{equation}
Since $P_{a'}$ is the identity on $\HC_{a'}$, $J'$ is an isometry with input
$\HC_{a'}$. In general, $J'$ maps its input to some subspace of
$\HC_{b} \ot \HC_{c}$. In the present case, this subspace is exactly $\HC_{b'}
\ot \HC_{c'}$.  
Thus, we write

\begin{equation}
    J' : \HC_{a'} \mapsto \HC_{b'} \ot \HC_{c'}.
\end{equation}
The above isometry generates a pair of quantum channels $\BC': \hat \HC_{a'}
\mapsto \hat \HC_{b'}$ and $\CC' : \hat \HC_{a'} \mapsto \hat \HC_{c'}$.
Since $\BC'$ is obtained by restricting the input space of $\BC$ to a
subspace, $\BC'$ is a subchannel of $\BC$.  Similarly, $\CC'$ a
subchannel of $\CC$.  

We shall be interested in the subchannels just defined to prove two facts.
\begin{lemma}
    The quantum capacity of $\CC_l$ satisfies
    \begin{equation}
        \log (d-1) \leq \QC(\CC_l).
    \end{equation}
\end{lemma}
\begin{proof}
    It follows from the definition of the quantum capacity in
    eq.~\eqref{chanCap} that the quantum capacity of a channel is bounded from
    below by the quantum capacity of any of its subchannels. Thus $\CC'$ is a
    subchannel of $\CC_l$, $\QC(\CC') \leq \QC(\CC_l)$. Since $\CC'$ perfectly
    maps its $d-1$ dimensional input $\HC_{a'}$ to its output $\HC_{c'}$,
    $\QC(\CC') = \log (d-1)$.  
\end{proof}
\begin{lemma}
    Channel $\BC_l$ is not degradable for any $0 \leq \lm \leq 1$.
\end{lemma}
\begin{proof}
It follows from the definition of degradability mentioned below
    eq.~\eqref{qpEqual} that if a channel is degradable then all its
    subchannels are also degradable.  This absence of degradability of $\BC$
    follows from an absence of degradability of the subchannel $\BC'$.  In fact
    one can show that $\BC'$ is anti-degradable, i.e., $\TC' \circ \CC' = \BC'$
    where $\TC'(A) = \Tr(A) [0]$. 
\end{proof}
\begin{lemma}
The channel $\BC_l$ is anti-degradable for $1/2 \leq \lm
\leq 1$, i.e.,

\begin{equation}
    \BC_l=  \EC \circ \CC_l,
    \label{ApBAdeg}
\end{equation}
for some quantum channel $\EC: \hat \HC_{cl} \mapsto \hat \HC_{b}$.
\end{lemma}
\begin{proof}
    To
construct $\EC$, consider an isometry $L:\HC_{cl} \mapsto \HC_{b} \ot
\HC_{e}$,

\begin{equation}
    L \ket{0} = \sqrt{1- \zt} \ket{00} + \sqrt{\zt/2}(\ket{11} +  \ket{22}), \quad
    L \ket{i} = \ket{0i},
\end{equation}
where $1 \leq i \leq d-1$, $\HC_e$ has dimension $d$, and $0 \leq \zt \leq
1$. Let $\EC$ be the channel generated by $L$ with channel input $\HC_{cl}$
and output $\HC_{b}$. Then for $1/2 \leq \lm \leq 1$, $\EC$ 
satisfies eq.~\eqref{ApBAdeg} with $\zt =(1-\lm)/\lm$.
\end{proof}

\subsection{Channel coherent information}
\label{AQutritInputCoh}

As discussed in the previous section, the isometry $J_l$ allows~\eqref{eq:iso2}
for leakage of quantum information to the environment.
Allowing such leakage, coupled with lack of any changes in the channel's output
space, is not expected to increase one's ability to pass noiseless quantum
information across noise described by the channel.  This conceptual expectation
is not without merit: next, we heuristically and numerically find allowing leakage
does not change the channel's coherent information $I$.

We argue for an equality,
\begin{equation}
    I(\BC_l) = I(\BC).
    \label{cohB1isB}
\end{equation}
The channel coherent information $I(\BC_l)$ is found by maximizing the
coherent information $\Dl(\BC_l, \rho)$ over all density operators $\rho$ on $\HC_{al}$. 
The $\HC_{al}$ space has a special $d-1$ dimensional subspace 
$\HC_{a'} = \text{span} \{ \ket{0}, \ket{i} \}$, $i \geq 2$.
All states in this subspace $\HC_{a'}$ are mapped perfectly to the channel environment
$\HC_{cl}$. Thus, a density operator $\rho$ with support in
$\HC_{a'}$ cannot help maximize $\Dl(\BC, \rho)$. However, if the support of
$\rho$ does not intersect with $\HC_{a'}$, then $\rho = \dya{1}$ and $\Dl(\BC_l,\rho) = 0$. 
A large, non-negative $\Dl(\BC_l,\rho)$ may be obtained only when the support of
$\rho$ is two dimensional, and this two-dimensional support has a one
dimensional intersection with $\HC_{a'}$. 
Such a two dimensional support must contain $\ket{1}$ and some $\ket{\psi}$ in
$\HC_{a'}$. To maximize $\Dl(\BC_l,\rho)$, choose $\ket{\psi}$ such that
$\BC([1])$ and $\BC([\psi])$ have small overlap $\Tr
\big(\BC([1])\BC([\psi])\big)$, while $\CC([1])$ and $\CC([\psi])$ have large
overlap $\Tr \big(\CC([1])\CC([\psi])\big)$. This can be arranged by letting
$\ket{\psi}$ be any linear combination of $\ket{0}$ and $\ket{2}$. All such
linear combinations are treated identically by $J_l$, thus a choice
$\ket{\psi}=\ket{0}$ can be made without loss of generality.
Thus $\Dl (\BC_l, \rho)$ is maximum when $\rho$ has support $\HC_{a}$.  Input
$\rho$ with support $\HC_{a}$ gives a channel output $\BC_l(\rho)$ which is
identical to $\BC(\rho)$, thus the maximum coherent information $\Dl(\BC_l,
\rho)$ is expected to equal the maximum of $\Dl(\BC, \rho)$. As a result, both
channels $\BC_l$ and $\BC$ are expected to have the same coherent information.

Numerics confirm the above heuristic. In these numerics with $d=3$,
$\Dl(\BC_l,\rho)$ is maximized by setting $\rho = A A^\dag/\Tr(AA^{\dag})$
where $A$ is an upper triangular matrix.  The matrix has $3$ real parameters on
the diagonal and $6$ real parameters representing the real and imaginary parts
of the off diagonal elements. To optimize the parameters, we use standard
numerical techniques for global optimization~(SciPy)~\cite{VirtanenGommersEA20}.

\section{Super-additivity of $I(\BC_l)$}
\label{sec:superAdd}

\subsection{A quick example}

Super-additivity allows random coding based error-correction across joint uses of quantum channels
to send noiselss quantum information at a higher rate than error-correction across
independent uses of those same channels.  Mathematically, given a channel $\BC$
the inequality $I(\BC^{\ot n}) \geq nI(\BC)$ can be strict for $n \geq 2$;
operationally, rates achievable asymptotically by coding in the typical
subspace of inputs to $\BC^{\ot n}$ can be larger than
$I(\BC)$~\cite{DiVincenzoShorEA98, SmithSmolin07, FernWhaley08,
LeditzkyLeungEA18, SiddhuGriffiths21, BauschLeditzky21}. 
This motivates us to consider an input to two uses of
$\BC_l$~(here and below forward we fix $d=3$ unless stated otherwise),
\begin{equation}
    \sg = (\dya{e_0} + \dya{e_1})/2,
\end{equation}
where
\begin{equation}
    \ket{e_0} = \ket{01}, \quad,
    \ket{e_1} = \sqrt{p} \ket{00} +  \sqrt{1-p}\ket{12}),
    \label{eq:input}
\end{equation}
and $0 \leq p \leq 1$ is chosen to be half. This entangled input maps to a separable output
$\BC_l^{\ot 2}(\sg)$ and an entangled environment state $\CC_l^{\ot
2}(\sg)$~(see App.~\ref{AQtritEnt}). 
At $\lm=1/3$ the coherent information of this state $\Dl (\BC_l^\ot 2, \sg) \simeq .067$ is positive
and thus noiseless quantum information can be sent across $\BC_l$ at a finite
rate, $I(\BC_l^{\ot 2})>0$, even though the $I(\BC_l)=0$ at $\lm = 1/3$~(see
eq.~\eqref{cohB1isB} and Fig.~\ref{FigE}).
Intuitively, the coherent information is non-zero because $\BC_l^{\ot 2}
(\dya{e_0})$ and $\BC_l^{\ot 2} (\dya{e_1})$ are well separated~(for instance
in $l_1$ distance~\cite{NoNamed}) in comparison to $\CC_l^{\ot 2} (\dya{e_0})$ and
$\CC_l^{\ot 2} (\dya{e_1})$.
In this next section we further explore when $I(\BC_l^{\ot 2})$ is non zero and when 
this quantity is super-additivity, $I(\BC_l^{\ot 2}) > 2
I(\BC_l^{\ot 2})$.

\subsection{Region where quantum capacity is (non)-zero}

Checking $\QC(\BC_l) > 0$ for all values $0 \leq \lm < 1/2$ can be
hard to resolve even numerically. However, we are able to resolve this using an
algebraic log-singularity based argument. Key is to apply the argument to two
copies of the channel $\BC_l$. 
\begin{theorem}
    For $d \geq 3 $ and $0 \leq \lm < 1/2$ 
    \begin{equation}
        0 < \frac{1}{2} I(\BC_l^{\ot 2}) \leq \QC(\BC_l).
        \label{eq:thineq}
    \end{equation}
\end{theorem}
\begin{proof}
Consider an input to $\BC_l \ot \BC_l$,
\begin{equation}
    \rho(\ep) := \ep \dya{e_0} + (1- \ep) \dya{e_1},
    \label{nonAddInp}
\end{equation}
where $\{\ket{e_0}, \ket{e_1}\}$ are defined in~\eqref{eq:input}, $0 < \ep
< 1$~(we suppress the dependence of $\rho(\ep)$ in~\eqref{nonAddInp} on $p$
for convenience).
Let $\Dl(\ep)$ denote the coherent information of $\BC_l^{\ot 2}$ at
    $\rho(\ep)$. A simple calculation shows that $\Dl(0) = 0$. In the interval
    $0 \leq \lm < 1/2$, $S_{bb}(\ep)$ and $S_{cc}(\ep)$ have
    $\ep\log$-singularities with rates 
\begin{equation}
    x_{bb} = 1 - \lm, \quad \text{and} \quad
    x_{cc} = \lm \big( 1+ \frac{p(1-\lm)}{2-2p(1-\lm)} \big),
    \label{lSingRate}
\end{equation}
respectively. The $\log$-singularity in $S_{bb}(\ep)$ is stronger, i.e.,
$x_{bb} > x_{cc}$ for all $0 \leq \lm < 1/2$ when $p$ is strictly less than
\begin{equation}
    p_{\max} := \frac{2(1-2\lm)}{(1-\lm )(2-3\lm)},
    \label{pMaxEq}
\end{equation}
where $p_{\max}>0$ for any $0 \leq \lm < 1/2$. This stronger singularity
implies that $\Dl(\ep) > 0$ and thus
\begin{equation}
    I(\BC^{\ot 2}_l) > 0,
    \label{Q2Pos}
\end{equation}
when $0 \leq \lm < 1/2$. This proves the first inequality in
eq.~\eqref{eq:thineq}.  The second inequality follows from the definition
of the quantum capacity.
\end{proof}

\begin{corollary}
    The quantum capacity $\QC(\BC_l)>0$ for $0 \leq \lm < 1/2$ and
    zero for $1/2 \leq \lm \leq 1$.
    \label{cr2}
\end{corollary}
\begin{proof}
    From the discussion in Sec.~\eqref{AQutritSub}, for $1/2 \leq \lm \leq 1$,
    $\BC_l$ is anti-degradable and thus $\QC(\BC_l) = 0$.  When $0 \leq \lm
    <1/2$, $\QC(\BC_l)>0$.
\end{proof}

\subsection{Two-letter level super-additvity}

Consider the family of density operators at the input
of $\BC_l \ot \BC_l$,
\begin{equation}
    \Lm = \sum_{i=0}^3 r_i \dya{e_i},
    \label{rhoOpt1}
\end{equation}
where $\{ \ket{e_0}, \ket{e_1} \}$ are defined in~\eqref{eq:input}, 
\begin{equation}
    \ket{e_2} = \ket{11}, \quad \ket{e_3} = \ket{10},
\end{equation}
%
and $r_i$ are real positive numbers that sum to unity.
Consider a parametrization where $r_3 = 1 - \sum_{i=0}^2 r_i$ is fixed while
$\rB = (r_0, r_1, r_2)$ and $p$ in~\eqref{eq:input} together constitute
four free parameters in $\Lm$.
Fixing these parameters gives some interesting special cases of $\Lm$. Let
$\Lm^{1}$ be a special case obtained at $p = 1$, $\rB = (z(1-z), z^2, (1-z)^2
)$ where $0 \leq z \leq 1$ is a free parameter. Let $\Lm^{2}$ be a special case
obtained by setting $\rB = (r_0, r_1, 1 - r_0 - r_1)$ and $\Lm^{3}$ be a
special case of $\Lm^{2}$ obtained by setting $r_0 + r_1 = 1$.  Notice
$\Lm^{1}, \Lm^2,$ and $\Lm^3$ have ranks four, three, and two respectively.
Let
\begin{equation}
    \Dl^* = \frac{1}{2} \underset{\{\rB, p\}}{\max} \;\Dl(\BC_l^{\ot 2}, \Lm).
    \label{eq:dlStarDef}
\end{equation}
The four parameter maximization to obtain $\Dl^*$~\eqref{eq:dlStarDef} is first
performed using standard maximization techniques~(SciPy). We find that for all
$\lm \leq \lm_0 \simeq .23$ the maximum obtained this way occurs at $\Lm^1$. As
$\lm$ in increased beyond $\lm_0$, $\Dl^*$ rises above $I(\BC_l)$ and the
maximizing density operator shifts from $\Lm^1$ to $\Lm^2$. As $\lm$ becomes
equal to $1/3$, $I(\BC_l)$ becomes zero and the maximizing density operator
changes from $\Lm^1$ to $\Lm^2$. The rank two density operator $\Lm^2$ has two
free parameters $r_0 = \ep$ and $p$. As $\lm$ is increased further and
approaches $1/2$, standard numerics become unstable. The origon of this
instability are logarithmic singularities that dominate the behavior of
$\Dl(\BC_l^{\ot 2},\Lm)$. Such numerical instability can be
tamed~(see App.~\ref{AQutritCoh2}) to obtain the maximum, $\Dl^*$.

\begin{figure}[ht]
    \centering
    \includegraphics[scale=.75]{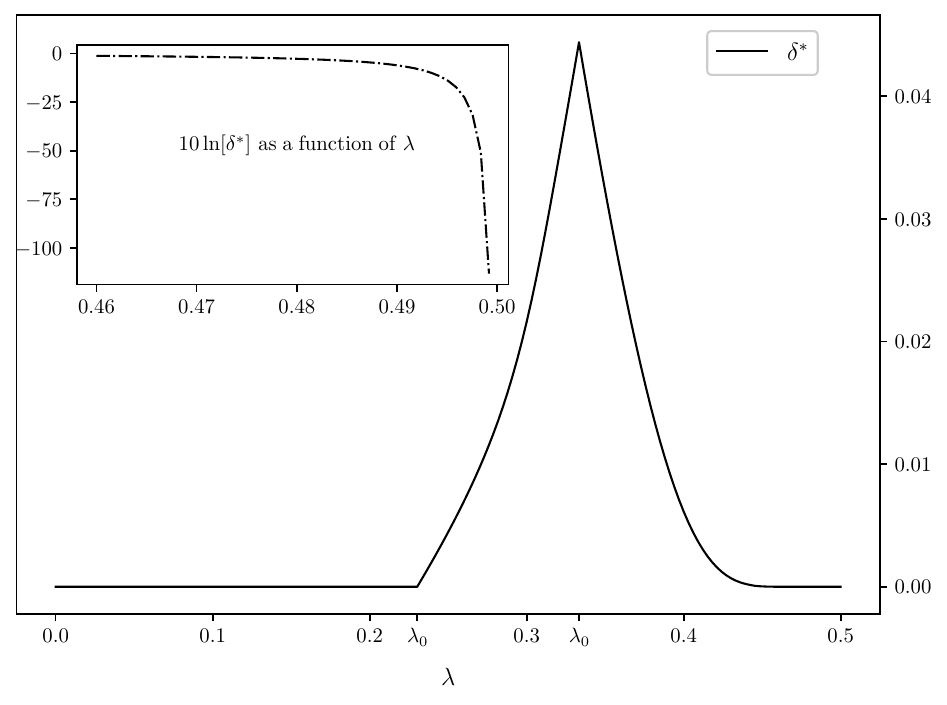}
    \caption{A plot of the difference $\dl^* = \Dl^*/2 - I(\BC_l)$. The
    inset shows $\dl^*$ on a logarithmic scale for $\lm$ close to $1/2$.}
    \label{FigF}
\end{figure}

\begin{figure}[ht]
    \centering
    \includegraphics[scale=.75]{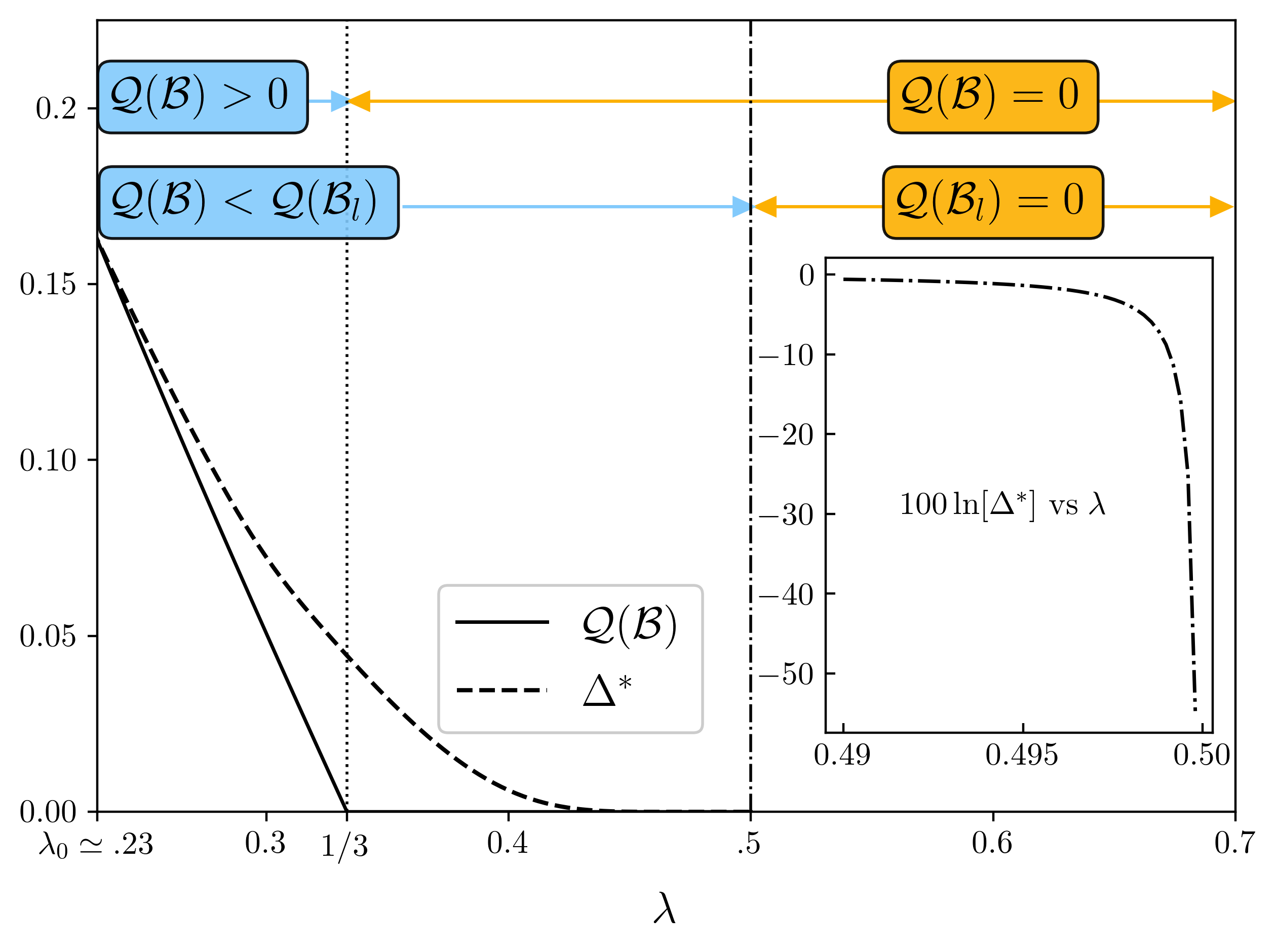}
    \caption{A plot of achievable rates for quantum communication across
    channels $\BC$ and $\BC_l$~(defined via~\eqref{eq:iso1} and~\eqref{eq:iso2}
    resp.).  As noise parameter $\lm$ is increased the quantum capacity of
    $\BC$~(thick curve) decreases to become zero at $\lm=1/3$. The lower
    bound $\Dl^*$~(dashed curve) on $\QC(\BC_l)$ is higher than $\QC(\BC)$
    for noise parameter values between $\lm_0 \simeq 0.23$ and $1/2$. This bound
    decreases with $\lm$, but using a careful analysis can be shown to remain
    non-zero for any $\lm < 1/2$~(see inset for numerical values of the bound
    in $\log$ scale). The bound falls to zero at $\lm = 1/2$ where $\QC(\BC_l) =
    0$.}
    \label{FigG}
\end{figure}

To understand the amount by which $\Dl^*$ exceeds $I(\BC_l)$ we plot,
\begin{equation}
    \dl^* := \Dl^* - I(\BC_l),
    \label{nonAddD}
\end{equation}
in Fig.~\ref{FigF}.  In this figure,
as $\lm$ in increased from zero, $\dl^*$ remains zero until $\lm$ reaches
$\lm_0 \simeq .23$, at which point $\dl^*$ becomes positive.  The value of
$\dl^*$ continues to increase and reaches a maximum $\simeq 4.4 \times 10^{-2}$
at $\lm = 1/3$, where $\dl^*$ has a sharp peak.
The peak is genuine and reflects the fact that $I(\BC_l)$ becomes zero at
$\lm = 1/3$ with finite slope while $\Dl^*$ is non-zero and strictly
decreasing.
As $\lm$ is increased beyond $1/3$, $\dl^*$ decreases.  For $\lm$ less than,
but in the vicinity of $\lm = 1/2$, $\dl^*$ is small but non-zero, as can be
seen from the values of $\dl^*$ plotted on a logarithmic scale in the inset
of Fig.~\ref{FigF}. At $\lm = 1/2$, $\dl^*$ becomes zero, $\BC$ becomes
anti-degradable and ceases to have quantum capacity. The region $\lm_0 \leq \lm < 1/2$ 
where $\dl^* >0$, the coherent information is super additive at the two-letter,
$I(\BC_l^\ot 2) > 2 I(\BC_l)$, by an amount $\dl^*$.

Our analysis of super-additivity relied on obtaining $I(\BC_l)$ by maximizing
$\Dl(\BC_l, \sg(z))$ over $\sg(z)$ of the special form in~\eqref{sgOpt} and
obtaining a lower boud on $I(\BC_l^{\ot 2})$ maximizing $\Dl(\BC^{\ot 2}_l,
\Lm)$ where $\Lm$ has the special form in~\eqref{rhoOpt1}.  While the special
form~\eqref{sgOpt} follows from arguments in Sec.~\ref{AQutritInputCoh},
additional numerics without using these two special forms did not lead to any
additional super-additivity.

\section{Relationship between capacity of extended and original channel}
\label{sec:QCapRel}

Our analysis thus far concerned the coherent information of $\BC_l$ over one
or several uses. It is natural to ask what can be said about the capacities
of $\BC_l$

We first note that the form of~\eqref{rhoOpt1} allows us to compare the quantum
capacity of $\BC$ and $\BC_l$.
\begin{lemma}
    The quantum capacity $\QC(\BC)$ bounds from below achievable rates across
    $\BC_l$ obtained from maximizing the coherent information over $\Lm$ ,
    i.e.,
    \begin{equation}
        \QC(\BC) \leq \Dl^* \leq \QC(\BC_l)
        \label{eq:lmNeq1}
    \end{equation}
\end{lemma}
\begin{proof}
    In Th.~\ref{th:QCapTh}, eq.~\eqref{lm3:eq1} implies that for each $\lm$
    there is a $\sg(z)= z \dya{0} + (1-z) \dya{1}$ such that $\Dl\big(
    \BC,\sg(z) \big) = \QC(\BC)$. Notice $\sg(z) \ot \sg(z)$ is equivalent
    to $\Lm^{1}$. Thus $\Dl^* = \Dl(\BC_l^{\ot 2}, \Lm)/2 = \QC(\BC)$
    at some fixed $\rB$ and $p$.  Since $\Dl(\BC_l^{\ot 2}, \Lm)$ at
    a fixed $\rB$ and $p$ is always smaller than the maximum value of
    $\Dl(\BC_l^{\ot 2}, \Lm)$ we obtain the first inequality in
    eq.~\eqref{eq:lmNeq1}.
    The second inequality in~\eqref{eq:lmNeq1} follows from eq.~\eqref{chanCoh}
    and~\eqref{chanCap}. These imply
    \begin{equation}
        \Dl^* \leq \frac{1}{2} I(\BC_l^{\ot 2}) \leq \QC(\BC_l).
    \end{equation}
\end{proof}

\subsection{Relationship between quantum capacity $\QC(\BC)$ and $\QC(\BC_l)$}

We find that for a range of channel parameter values $\lm_0 \simeq .23$ to
$1/2$, the channel $\BC_l$, which allows for
leakages, can send more quantum information than $\BC$, which doesn't allow for
leakage~(see Fig~\ref{FigG}). We show this claim in two parts, the first is an
algebraic proof for the parameter range $1/3 \leq \lm < 1/2$ and the second is
numerical where $1/3$ is replaced with $\lm_0 \simeq .23$.

\begin{theorem}
    \label{th:mainTh}
    The capacity of channels $\BC$ and $\BC_l$ satisfy an inequality
    \begin{equation}
        \QC(\BC) \leq \QC(\BC_l)
    \end{equation}
    which is strict for $1/3 \leq \lm < 1/2$ and is an equality for $1/2 \leq
    \lm \leq 1$.
\end{theorem}
\begin{proof}
    The inequality is obvious since $\BC$ is a sub-channel of $\BC_l$ obtained
    by restricting the input, $\HC_{al}$, of $\BC_l$ to a subspace $\HC_a$.
    To prove the inequality is strict for $1/3 \leq \lm < 1/2$ notice that for
    $1/3 \leq \lm \leq 1$, $\BC$ is anti-degradable~(see
    Lemma~\ref{lem:degADeg}) and thus $\QC(\BC) = 0$ however $\QC(\BC_l)>0$ for
    all $0 \leq \lm \leq 1/2$~(see Corollary~\eqref{cr2}).
\end{proof}

Numerics indicate that the inequality in Th.~\ref{th:mainTh} is strict
for a larger range of $\lm$ than stated in Th.~\ref{th:mainTh}. In particular,
it follows from the discussion below eq.~\eqref{nonAddD} and ~\eqref{cohB1isB}
that $\QC(\BC_l) \geq \Dl^{*}>\QC(\BC)$ for $\lm_0 \leq \lm < 1/3$.

In addition to the quantum capacity, we compare the private classical capacity
of $\BC$ and $\BC_l$.

\begin{corollary}
    The private capacity of $\BC_l$ is larger than that of $\BC$, in
    particular,
    \begin{equation}
        \QC(\BC) = \PC(\BC) \leq \QC(\BC_l) \leq \PC(\BC_l)
    \end{equation}
    where the first inequality is strict for $1/3 \leq \lm \leq 1/2$.
\end{corollary}
\begin{proof}
    Proof for the first equality follows from the fact that $\BC$ is either
    degradable or anti-degradable~(see Lemma~\ref{lem:degADeg}), and thus the
    quantum and private capacities of $\BC$ are equal~(see below
    eq.~\eqref{qpEqual}). The first inequality is a restatement of
    Th.~\ref{th:mainTh}, while the final inequality is the well-known
    relationship between a channel's quantum and private capacity~(recall
    discussion below eq.~\eqref{pInfDef}).
\end{proof}

In the interval of values $\lm_0 \leq \lm < 1/3$, where $\dl^*$ is positive,
the quantum and private capacities of $\BC_l$ are strictly larger than the
quantum and private capacities of $\BC$. In light of discussions containing
eqns.~\eqref{nonAddD}, the amount by which $\QC(\BC_l) \leq
\PC(\BC_l)$ is larger than $I(\BC) = \QC(\BC) = \PC(\BC)$ is at least $\dl^*$.
Thus $\dl^*$ indicates a magnitude of boost in capacities $\QC$ and $\PC$,
found by transforming $\BC$ to $\BC_l$. Notice the boost occurs over an interval
which includes $1/3 \leq \lm < 1/2$, where $\QC(\BC) = \PC(\BC)= 0$. Thus,
the boost not only increases the magnitude of the quantum and private
capacities of $\BC$, but it also extends the parameter interval over which
these capacities are non-zero. 

\section{Discussion}
\label{sec:disc}

Super-additivity is responsible for restricting our understanding of channel
capacities~\cite{Hastings09, LiWinterEA09, SmithSmolin09}. For instance a
channel's quantum capacity can only be written as an infinite
limit~\cite{Shor02a, Devetak05}. Showing that the capacity is non-zero can
require computation of $I(\BC^{\ot n})$ over an unbounded number of channel
uses $n$~\cite{CubittElkoussEA15, ElkoussStrelchuk15}. Beyond these fundamental
difficulties, in practice a channel's quantum capacity $\QC(\BC)$ can be
strictly larger than its coherent information $I$. Over the past 25 years the
amount by which $\QC$ can exceed $I$ in physically motivated channels has been
pushed to no more than $O(10^{-3})$~\cite{SmithSmolin07, LeditzkyLeungEA18,
Filippov21, SiddhuGriffiths21}. It would be consistent with prior work to hold
the view that super-additivity is a small effect.
In this paper we find that super-additivity can be an order of magnitude
larger.  At $\lm \simeq 1/3$, we find $\QC(\BC_l)$ exceeds $I(\BC_l)$ by at
least $O(10^{-2}) \log_2 3$~(the $\log_2 3$ factor normalizes by the capacity
of an identity channel of the same input dimension as $\BC_l$).  This excess is
found using a simple but valuable input~\eqref{rhoOpt1}.

While having a large amount of super-additivity boosts rate, the noise level at
which super-additivity begins also matters. It is known that in the limit of
very low noise, super-additivity must vanish~\cite{BrandaoOppenheimEA12, 
LeditzkyLeungEA18a}. So far, 
superadditivity has been found in regions of high noise: either $I = 0$
or $I/\log d_a \simeq O(10^{-2})$~\cite{FernWhaley08, BauschLeditzky21}.
If super-additivity only occurs with high noise, then perhaps in some practical
channels with moderate or low-noise it may be ignored. We argue it may be
sub-optimal to ignore these low-noise regions, i.e., super-additivity can also
appear in moderate to low noise setting. We find $\QC > I$ starting at $\lm$
values where $\QC/\log_2 3 \simeq 10 \%$.
Super-additivity has been thoroughly explored in the frontier of high noise
where a channel $\BC$'s coherent information is zero but the quantum capacity
is not known to be zero.
The explorations seek to answer a fundamental question in quantum information,
given a channel can it send quantum information?
Explorations on various physically motivated low-dimensional channels using a
variety of approaches based on machine learning~\cite{BauschLeditzky20},
non-convex optimization~\cite{FernWhaley08, SidhardhAlimuddinEA22}, and
degenerate quantum codes~\cite{SmithSmolin07} have been done but success has
been partial. There has always been a noise parameter of the channel for which
$I = 0$, the best known lower bound on capacity is also zero~(no
super-additivity is found) but the quantum capacity isn't known to be zero.
One reason for this lack of success may be that it may take unbounded number of
channel uses $n$ to find super-additivity. 
A key feature of the super-additivity discussed here is that it occurs for just
two uses of $\BC_l$ and it occurs in a maximal way, i.e., it occurs everywhere
$I(\BC_l) = 0$ but the capacity can be non-zero.
The coherent information becomes zero when the noise parameter $\lm = 1/3$, the
quantum capacity becomes zero only at $\lm = 1/2$. For every
parameter in the wide range, $1/3 < \lm < 1/2$, we dramatically find
super-additivity at the two-letter level, $I(\BC_l^{\ot 2})>0$.  In
contrast to numerical methods, the finding, $I(\BC_l^{\ot 2})>0$, is
purely algebraic.

Sending classical communication hidden from a channel's environment is among
the most significant applications of quantum
information~\cite{BennettBrassard14}. However, an ultimate rate for carrying
out such communication, the private capacity $\PC$, is hard to calculate and
fully understand~\cite{Devetak05, CaiWinterEA04}. We find that allowing leakage
of information to the environment can boost $\PC$.  Since quantum communication
is necessarily private, $\PC \geq \QC$, however since $\BC$ is degradable
$\QC(\BC) = \PC(\BC)$.  As $\BC$ is modified to $\BC_l$ by allowing leakage,
the quantum capacity increases, i.e. $\QC(\BC) < \QC(\BC_l)$ and as a result
the private capacity $\QC(\BC) = \PC(\BC) < \QC(\BC_l) \leq \PC(\BC_l)$ also
increases.  Since modification of $\BC$ to $\BC_l$ accompanies a loss of
degradability it is not clear if this last inequality is an equality.  It would
be interesting to study if the private capacity of $\BC_l$ differs from its
quantum capacity.  Such separations can occur~\cite{HorodeckiHorodeckiEA05,
HorodeckiPankowskiEA08, WangFangEA18, LeditzkyLeungEA23}, and have led to
non-trivial insights about quantum information and key
distribution~\cite{SmithYard08, LeditzkyLeungEA23a}. However, finding when and
why such separations occur is not trivial~\cite{LeditzkyLeungEA18}. 
We leave this, and other such investigations of how our leakage based
transformation may effect the two-way quantum, the classical and various other
capacities of quantum channels~\cite{BennettDiVincenzoEA96, Holevo20}, as an
interesting area of future study.

Super-additivity reported here is exciting for theoretical, experimental, and
practical reasons.  Strong theoretical efforts are being made to design
error-correcting codes that correct noise from channels with zero coherent
information, this extended reach in error correction takes advantage of
super-additivity~\cite{BonillaAtaidesTuckettEA21, HuangPesahEA23,
RoffeCohenEA23}.
However, when super-additivity occurs over a narrow range of parameters, this
reach can be limited, less robust to noise and thus hard to confirm using
numerics and algebra.  At present quantum codes seek this advantage by testing
their performance with Pauli channels. However, the extent of super-additivity
in Pauli channels is much narrower than the maximal extent reported here.
It would be interesting to explore how the large amount of
super-additivity found here can assist ongoing efforts to design and test
error-correcting codes which seek to leverage super-additivity.
In addition to theoretical efforts, experimental efforts are being made to
observe super-additivity in some way~\cite{YuMengEA20}. Such observations are
easier to conduct when the super-additivity to be observed has a large
magnitude, wide extent, and appears over small channel uses. The
super-additivity reported here is much larger, has an extremely wide extent and
it occurs for the smallest number of channel uses, two. These features of
super-additivity in the simple qutrit channel introduced here make this channel
a promising experimental and theoretical test-bed for boosting current efforts
of exploring and understanding super-additivity.

\section*{Acknowledgments}
We thank Robert B. Griffiths, Sridhar Tayur, Graeme Smith, and John Smolin
for their support and valuable discussions, and Felix Leditzky for many useful 
comments. This work was partially supported by NSF CAREER Award CCF
1652560 and NSF Grant PHY 1915407

\appendix

\section{Asymptotic estimates for lower bound on $\QC(\BC_l)$}
\label{AQutritCoh2}
Consider the special input $\rho(\ep)$ in eq.~\eqref{nonAddInp}, it has two
parameters to $0 < \ep < 1$ and $0<p<1$.
As $\lm$ approaches $1/2$, i.e., for small positive $\dl \lm := 1/2 - \lm$, we
find that the maximum value of $\Dl(\BC^{\ot 2}_l, \rho(\ep))$ occurs at $\ep$
tending to zero. In this small $\dl \lm$ limit, standard numerical maximization
becomes unstable because one is trying to maximize a function of the form,

\begin{equation}
    f(p,\ep) = -\al \ep \ln \ep + \bt \ep,
    \label{fMax}
\end{equation}
where $\ln$ is the natural logarithm, $\al = (x_{bb} - x_{cc})/\ln 2$ is the
difference of $\ep \log$ singularity rates~\eqref{lSingRate} divided by $\ln
2$, and $\bt$ depends on $p$ and $\lm$.  If $\al$ is positive, then $f$ is
positive for small enough $\ep$.  From comments above eq.~\eqref{pMaxEq}, it
follows that as long as $0 < p < p_{\max}$, $\al$ is positive for any small
positive $\dl \lm$. One may express $\al$ and $\bt$ as follows

\begin{equation}
    \al = \al_0 \dl \lm + \al_1 (\dl \lm)^2 + \dots, \quad 
    \bt = \bt_0 + \bt_1 (\ln \dl \lm) + \bt_2 (\dl \lm \ln \dl \lm) + \bt_3 (\dl \lm) + \dots,
\end{equation}
where $\al_0 = 2(1-r)/\ln 2$, $\al_1 = 8 r \al_0$, $\bt_0 = 1+ \ln r/(4 \ln 2)$,
$\bt_1 = 1/(4 \ln 2)$, $\bt_2 = (2r + 1/2)/\ln 2$,

\begin{equation}
    \bt_3 = \frac{1}{\ln 2}[-3/2 - 9r \ln 2 + 8r \ln(16 r) + \ln (256 r)/2],
\end{equation}
and $r:= p/p_{max}$. For small $\dl \lm$, we use the above expressions
in eq.~\eqref{fMax} and maximize $f(p, \ep)$ to obtain $\Dl^*$. For this
maximization it is convenient to replace $p$ with $r$ where $r$ lies between
zero and one. The replacement gives a function $g(r,\ep)$ which equals $f(r
p_{\max}, \ep)$. For small $\dl \lm$, the values of $g$ are also small, hence
one maximizes $\ln g$. To obtain this maximum, we set to zero the derivative of
$g(r,\ep)$ with respect to $\ep$. This occurs at

\begin{equation}
    \ep = \ep^* := \exp[ -( 1- \bt / \al)].
\end{equation}
Next, we numerically maximize 

\begin{equation}
    \ln g(\ep^*,r) = \ln (\al) -(1- \bt/\al),
\end{equation}
over $0<r<1$ to obtain $\ln \Dl^*$. The values of $\ln \Dl^*$ obtained this way
are in good agreement with standard numerical calculations for moderate values
of $\dl \lm$. In the inset of Fig.~\ref{FigG} in the main text, we plot $\ln
\Dl^*/2$ for small $\dl \lm$.
Since $I(\BC_l) = 0$ for small $\dl \lm$, $\ln \Dl^*$ readily gives $\ln
\dl^* = \ln[\Dl^* - I(\BC_l)] = \ln[\Dl^*]$.

\section{Entanglement in input and outputs}
\label{AQtritEnt}
A bipartite density operator $\rho_{aa}$ on $\HC_{a} \ot \HC_{a}$ is separable
if it can be written as

\begin{equation}
    \rho_{aa} = \sum_i p_i [\al_i] \ot [\al'_i],
\end{equation}
where $p_i$ are positive numbers that sum to one, $\ket{\al_i}$ and
$\ket{\al'_i}$ are pure states in the first and second $\HC_a$ spaces,
respectively. The density operator $\rho_{aa}$ is entangled across $\HC_a \ot
\HC_{a}$ if it is not separable. The PPT condition~\cite{Peres96} for
entanglement states that $\rho_{aa}$ is entangled if its partial transpose with
respect to the second $\HC_a$ space gives an operator with negative eigenvalues.

In the main text we mentioned that $\sg = (\dya{e_0} + \dya{e_1})/2$ and
$\sg_{cc} = \CC_l^{\ot 2}(\sg)$ are entangled while $\sg_{bb} = \BC_l^{\ot
2}(\sg)$ is not. The partial transpose of $\sg_{aa}$ with respect to the second
$\HC_a$ space gives an operator with one negative eigenvalue -1/4, and the
partial transpose of $\sg_{cc}$ with respect to the second $\HC_c$ space gives
an operator with one negative eigenvalue $(3-\sqrt{21})/24$. Thus the PPT test
shows that both $\sg$ and $\sg_{cc}$ are entangled. It is easy to check
that $\sg_{bb}$ is diagonal in the standard basis $\ket{i} \ot \ket{j}$, thus
it is separable.


\begin{thebibliography}{10}
\bibitem{NoNamed}
$l_1$ distance between two density operators $\rho_a$ and $\sg_a$ is
  $\Tr(\sqrt{N^{\dag}N})$ where $N = \rho_a - \sg_a$.
\end{thebibliography}
\end{document}